\documentclass[aps,prd,twocolumn,showpacs,superscriptaddress,groupedaddress]{revtex4-1}
\usepackage{graphicx}  
\graphicspath{ {./graphics} }
\usepackage{comment, hyperref}
\usepackage{bm}        
\usepackage{amssymb}

\begin{document}

\title{Exploring viscous damping in undergraduate Physics laboratory using electromagnetically coupled oscillators}

\author{N. Jayaprasad}
\author{P. Sadani}
\author{M. Bhalerao}
\affiliation{B. Tech. Sophomore Year, IIT Gandhinagar, Gujarat, India.}

\author{A.~S.~Sengupta} \email[email: ]{asengupta@iitgn.ac.in} 
\author{B.~Majumder} \email[email: ]{barunbasanta@iitgn.ac.in} 
\affiliation{Department of Physics, IIT Gandhinagar, Gujarat, India.}

\date{\today}

\begin{abstract}
	We design a  low-cost, electromagnetically coupled, simple harmonic  oscillator and demonstrate free, damped and forced oscillations in an under-graduate (UG) Physics laboratory. It consists of a spring-magnet system that can oscillate inside a cylinder around which copper coils are wound. Such demonstrations can compliment the traditional way in which a Waves \& Oscillations course is taught and offers a richer pedagogical experience for students. We also show that with minimal modifications, it can be used to probe the magnitude of viscous damping forces in liquids by analyzing the oscillations of an immersed magnet. Finally, we propose some student activities to explore non-linear damping effects and their characterization using this apparatus.
\end{abstract}


\maketitle

\section{Introduction}

According to Farady's law, a magnet in relative motion to a surrounding coil, produces an induced voltage (emf) across it, whose magnitude can be shown to be proportional to the instantaneous velocity of the magnet. This finds widespread practical applications: Mossbauer spectrometer drives based on this principle are well described in \cite{waheed}. The same physical principle forms the theoretical basis for the design of inductive sensors to measure high-frequency current pulses \cite{Rojas}. An interesting illustration of how emf generated due to oscillation of one set of spring-magnet-coil systems can drive oscillations in another such system can be found in \cite{manos}. The study of such systems is essential to verify the right hand rules of electrodynamics.

The idea of the experiment presented in this paper, is to study free and forced oscillations and use it for demonstrating various concepts in waves and oscillations \cite{Ouseph}. Lecture demonstrations play a crucial role in bridging the gap between theoretical instruction and conceptual learning \cite{Miller} - and we believe such low-cost apparatus can be useful. 

The apparatus consists of a spring magnet system designed to have low natural frequency (about 1.5 Hz), that is suspended from a rigid frame in a plastic cylindrical beaker surrounded by two sets of coaxial coils. Free oscillations of the magnet result in the generation of a time-varying (sinusoidal) emf across one of the coils that can be viewed on a storage oscilloscope. The system energy can be dissipated by shorting the other set of coils - thereby closing the circuit and inducing resistive losses. In this case, the voltage induced across the pickup coil shows an exponentially damped sinusoidal time series. This may be recorded and analyzed by students to find a connection between the damping coefficient and the resistance in the coils. Further, one can drive this system by connecting one set of coils to a AC function generator at arbitrary frequencies and amplitudes. Students can record the amplitude response of the oscillator at different frequencies and graphically represent it to visualize the Lorentzian resonance curve, and extract relevant physical parameters from  observations. Some of the subtler features of the theory of oscillations can also be observed from  the transient response of the oscillator at early times. This enables a richer pedagogical experience in a UG laboratory in teaching such concepts and often compliments a mathematical treatment. 

\begin{figure*}
	\includegraphics[width=0.75\textwidth]{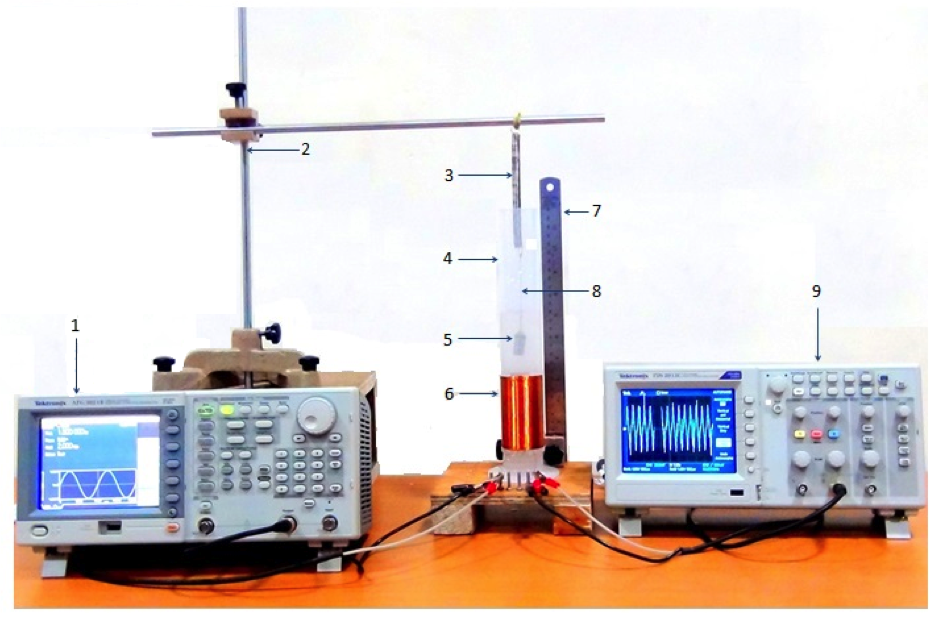}
	\caption{Details of the electromagnetically coupled oscillator apparatus: 1. function Generator, 2. rigid metal stand, 3. spring, 4. cylindrical beaker, 5. magnet, 6. coils, 7. ruler, 8. thin metal wire, 9. storage oscilloscope.}
	\label{fig:apparatus}
\end{figure*}

We show that such a simple device can also be useful in introducing the concepts of viscous damping, wherein the magnet is immersed in a viscous medium and sustains drag forces that are  proportional to its instantaneous velocity (to leading order). We demonstrate two different ways which couple the concept of viscous damping to those of free oscillations and discuss ways how students might characterize the strength of such damping forces in a viscous liquid. We also present few ideas for students to extend this to the realm of non-linear damping as well. 

The paper is organized as follows: in Section \ref{sec:apparatus}, we briefly review the design and construction of the apparatus. Some student activities are also proposed. In Section \ref{sec:theory}, we briefly review the theory of linearly damped oscillators and highlight the model amplitude response in such systems. We then describe a set of laboratory activities to characterize the magnitude of viscous damping forces. We also discuss how this may be related to the coefficient of viscosity in idealized conditions. Finally, we present our conclusions along with some suggestions for exploring the system further to characterize higher order effects of viscous damping.

\section{Apparatus and Setup}
\label{sec:apparatus}

The apparatus consists of a stainless steel spring (force constant $k = 1.39$ N/m) with a magnet (mass $m = 12 $ gm) attached at one end. The spring constant is measured by a Hooke's law apparatus assuming linear response $(k x = F_s)$ of the spring under a load $F_s$. The natural frequency of this system is $\omega_0 = \sqrt{ k/m }$ and is designed to be large enough so that adequate number of cycles can be recorded for analysis without too much disturbance in the liquid (when immersed). 

The length of the elongated spring with the magnet is 28 cm. The maximum displacement of the magnet is limited to 7 cm, as higher amplitudes may also produce undesirable turbulence when the magnet is immersed in liquid. The magnets are cylindrical, each of length 2 cm and diameter 1 cm. 

The spring-magnet is suspended from a rigid frame and lowered into a plastic cylindrical beaker (as shown in Fig \ref{fig:apparatus}) of length 19 cm, with inner and outer diameters of 4.2 cm and 4.5 cm respectively. The diameter of the cylinder used is about 4 times the dimension of the magnet to ensure that the latter's motion is not hindered by the cylinder during oscillations. Two separate coils of 500 turns each, made of $\#32$ gauge enameled copper wire are wound over a length of 8.3 cm around the cylinder. The resistance of each coil is $23\; \Omega$.

One of the coils called the `pick-up coil' detects the induced emf generated which can be viewed on a storage oscilloscope (model: Tektronix TDS 2012C). The induced voltage across the coil is directly proportional to the instantaneous velocity of the magnet. For observing forced oscillations, the other coil (called the `driving coil') is connected to a arbitrary function generator (model: Tektronix AFG3021B) which drives the system near its resonant frequency. It is found that 2V (peak-peak) voltage generated by the function generator is sufficient to obtain a significant resonant amplitude response of the system. The driving coils can also be shorted as needed to study electromagnetically damped oscillations. In the latter part, while characterizing linear viscous damping forces with this apparatus, the beaker is filled with water to a height of 15 cm. As the compression of the spring also gets affected inside the liquid, the magnet is lowered into water by a thin stainless steel wire.  

Since the induced emf is proportional to the velocity of the magnet (and not its displacement), the oscilloscope trace cannot be directly used to measure amplitude response. To mitigate this problem, a measuring scale is mounted behind the cylinder for measuring the amplitude of the oscillating magnet. For accurate measurements, the motion of the magnet is video recorded and played back in slow motion to record the readings. 

\begin{figure*}
	\includegraphics[width=0.495\textwidth]{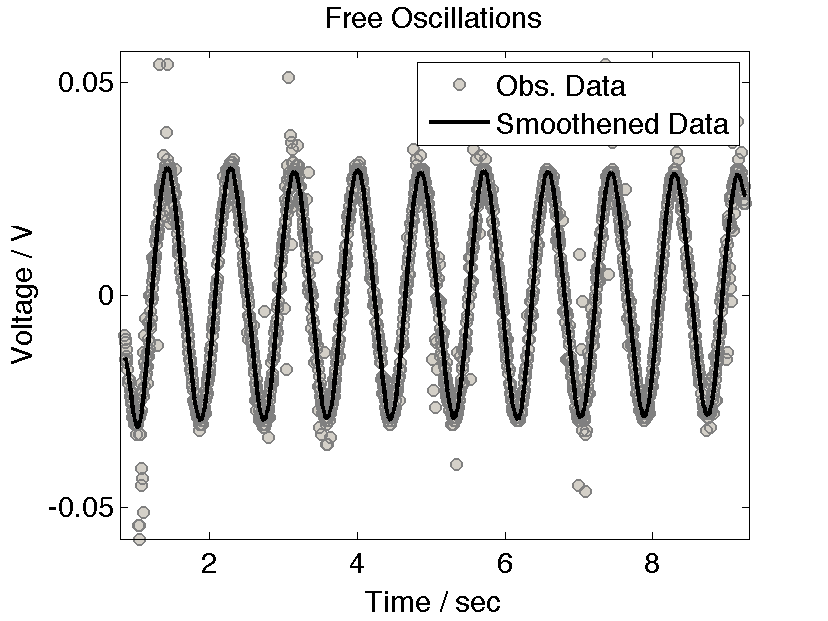}
	\includegraphics[width=0.495\textwidth]{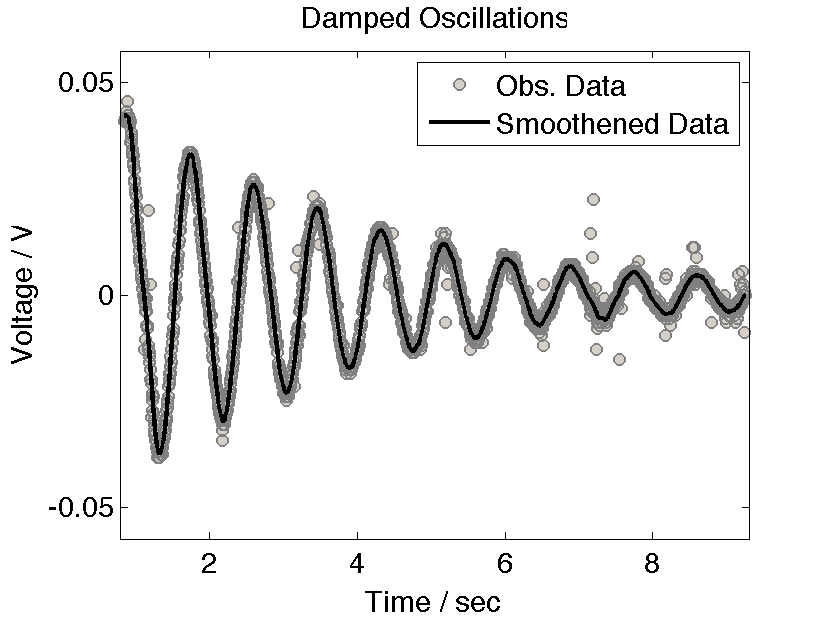}
	\caption{Free and electromagnetically damped oscillations as observed in the laboratory. The motion of the magnet through the coils produces an emf across the 'pickup' coil that is proportional to its instantaneous velocity. When the other set of coils is shorted, a current flows through it and the energy of the system is lost via resistive losses. The direction of this current $I$ is such that in any element $d\vec \ell$ of the coil, a resistive force $I \; d\vec \ell \times \vec B$ opposes the motion of the magnet. This acts like a linear damping term due to which the system steadily loses its energy (seen as a exponentially decreasing amplitude) on the right panel.}
	\label{fig:exampleOsc}
\end{figure*}

This apparatus provides an excellent means of studying free, damped and forced oscillations to augment theoretical instructions in a UG Waves \& Oscillations course. For example, students can be asked to record the induced voltage in the 'pickup coil' when the magnet oscillates freely in the cylinder. In this case there is no loss in the energy of the system as there is negligible drag due to air, and any initial displacement of the magnet is sustained unimpeded for a very long time. Students can then be asked to short the other set of coils and note the change in the induced emf pattern. In this case, current flows through the shorted coils of finite resistance and impedance leading to energy losses of the system over time. This is manifested as a steady decrease in the magnet's velocity and can be seen as a exponentially damped sinusoid in the storage oscilloscope. A related exercise would be to theoretically explain how the current in the shorted circuit leads to a damping term proportional to the velocity of the magnet as worked out in \cite{McCarthy}. This leads to an appreciation of the resistive and inductive nature of the shorted coils. A variable resistance, capacitance and inductance (RLC) box connected to the shorted coil can be used as a playground for the class to explore the system by changing  various parameters. The two cases above are highlighted in Fig. \ref{fig:exampleOsc}. A related exercise in the damped oscillation case would be to record the envelope of the decaying amplitude as a function of time and (a) checking that it indeed falls off as a an exponential by plotting it appropriately and (b) estimating the {\it effective} resistance in the shorted coils by measuring the decay constant. This would also help them visualize the solutions of a damped harmonic oscillator.

When the magnet is immersed in water, it's motion is opposed by the forces due to dynamic viscosity of the liquid. The magnitude of this viscous force $F_v$ is proportional to its instantaneous velocity (to linear order) and is given by the equation $F_v = -b v$. In order to estimate linear damping by the viscous drag alone, it is important to make sure that the coils are not shorted. Under idealizations of laminar flow and small values of $v$, this proportionality factor is linearly related to the medium's coefficient of viscosity $\eta$ and a geometrical factor  $\kappa_g$ \cite{Leith} that depends on the shape of the immersed magnet: $b = 3\pi \kappa_g \eta$. Estimating the magnitude of resistive forces in the viscous liquid essentially boils down to estimating the numerical value of $b$.  It is not our aim to measure the coefficient of viscosity using this apparatus: $\eta$ is conventionally measured by a viscometer \cite{visco, pipe} which is based on Hagen-Poiseuille law.

\section{Concise review of damped oscillations}
\label{sec:theory}

In this section we briefly outline the theory of linearly damped oscillations. This will serve to set up model amplitude responses against which we can compare our observations later. 

Let a spring-magnet system (of mass $m$ and natural frequency $\omega_0 = \sqrt{k/m}$) be set oscillating in a viscous liquid of damping coefficient $b > 0$. Further, let this system be driven by an external periodic driving force. From Newton's second law of motion, the instantaneous acceleration $\ddot x$ on the magnet can be related to the net force acting on it - which is a sum of the applied periodic force, the restoring force ($-k x$) and the resistive damping force ($-b \dot x$). This can written as:   
\begin{equation}
	\ddot{x} + 2 \beta \dot{x} + \omega_0^2 x = \frac{{\mathcal{F}}_0}{m} \sin (\omega t)~,
	\label{eq:Master}
\end{equation}
where ${\mathcal{F}}_0 \sin (\omega t)$ is the external force driving the system at a frequency $\omega$ and $\beta = b/2m$. The solution to this second order ordinary differential equation (ODE) is given by
\begin{equation}
	x(t) = A_0 \; e^{-\beta t} \cos (\omega t ) + \frac{F_0}{\sqrt{(\omega^2 - \omega_0^2)^2 + 4\beta^2 \omega^2}} \cos(\omega t + \phi)
	\label{eq:solMaster}
\end{equation}
where $A_0$ is an arbitrary constant (to be determined from initial conditions), $F_0 = {\mathcal{F}}_0/m$ and $\phi = \tan^{-1} [ 2\beta \omega / (\omega^2 - \omega_0^2)]$ is the phase of oscillations \cite{french}. This solution consists of two parts: the first term represents the transient solution which decays with time depending on $\beta$, while the second term gives the steady state (late time) oscillations (which becomes dominant after the transients decays out). In steady-state, the amplitude response is given by
\begin{equation}
	\label{eq:drivenOsc}
	\vert A (\omega) \vert = \frac{F_0}{\sqrt{(\omega^2 - \omega_0^2)^2 + 4\beta^2 \omega^2}} ~.
\end{equation}

In absence of a driving force, the right hand side of Equation (\ref{eq:Master}) can be set to zero, which then admits the damping solution
\begin{equation}
	x = A_0 \; e^{-\beta t}\cos \left [ (\omega_0^2 - \beta^2) t \right ]
	\label{eq:solDamped}
\end{equation}
Here, the envelope of amplitude decays exponentially as $e^{-\beta t}$. Note that the induced emf (observed on the oscilloscope) is proportional to $\dot x$, and will {\em also} have the same a similar envelope.

The above solution is valid when the damping force is linearly proportional to velocity. Additional damping forces proportional to the square of the velocity can be incorporated by adding a term $\left [ {\mathrm{sgn}}(\dot x) \gamma \dot x ^2 \right ]$ (where $\gamma > 0$ is a dimensionful constant) to the left hand side of Equation (\ref{eq:Master}). Without the ${\mathrm{sgn}}(\dot x)$ factor, this term would act as a damping term in one half cycle (when $\dot x$ is positive) and as a source of (spurious) energy in the next half cycle where the velocity changes sign. Any non-linear term with even power in velocity will need to have this factor to properly account for damping. However such a ODE is difficult to solve analytically without recourse to simplifying assumptions. From the above discussion, it should be clear that a damped oscillator is characterized by only two parameters: the natural frequency ($\omega_0$) and the damping coefficient ($\beta$).

\subsection*{Estimation of viscous forces in water}
\label{subsec:dataAnalysis}

With the theoretical understanding of damped oscillations above, we now proceed to estimate the magnitude of $\beta$ which is a measure of linear damping forces in water. 

\begin{figure}
	\includegraphics[width=0.95\columnwidth]{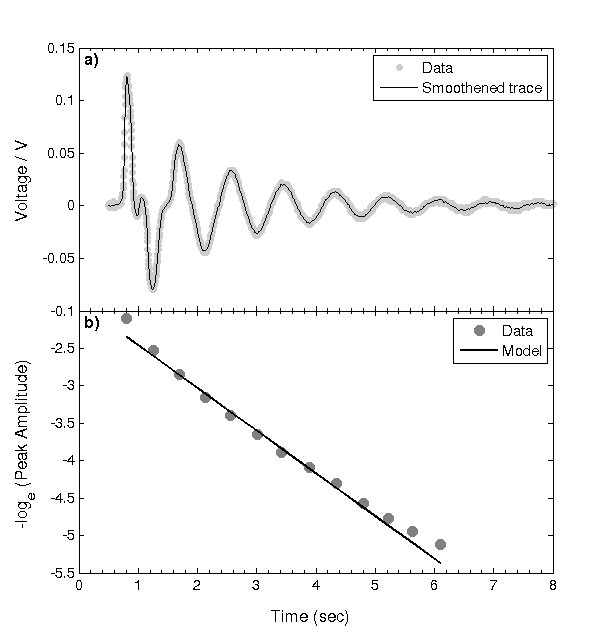}
	\caption{(a) Showing the damping of the freely oscillating magnet's velocity when immersed in water. In this case, the viscous forces in water offer a drag force that is linear in the magnet's velocity (to leading order) which leads to a dissipative loss in the system's energy. (b) Showing the peak of the amplitude (envelope of the trace in (a) above) falling off exponentially. This is expected for linearly damped oscillators from theory. When plotted in logarithmic scale, a straight line fit best describes the data. The (negative) slope of the line is the measure of $\beta$ as explained in the text.}
	\label{fig:dampingInWater} 
\end{figure}

A trivial way to estimate this quantity is to let the magnet freely oscillate in the beaker filled with water after an initial impulse and record the induced emf in time. By {\em freely} we mean that none of the coils surrounding the beaker are shorted or connected to the function generator. The total energy of the system is dissipated leading to loss in amplitude as seen in Figure \ref{fig:dampingInWater}(a). From theory, we expect the amplitude to decay exponentially with time as $\exp (-\beta t)$. As discussed earlier in the context of electromagnetic damping, one plots the natural logarithm of the amplitude peaks in each cycle vs time and expects the plot to be a straight line as seen in Figure \ref{fig:dampingInWater}(b). The (negative) slope of this line determined by the method of linear regression analysis gives an estimate of $\beta = 0.571$ directly, with 95\% confidence interval (CI) of $[0.523, 0.619]$. If the quality factor of the system is small, we can only observe a few cycles before the amplitude decays substantially and meaningful values of the peaks cannot be extracted from the data. In our case, we are able to estimate $\beta$ using about 6-7 cycles. However, one must be careful to ignore the first few cycles to discount for any transient effects from the initial impulse given to the magnet. 

Yet another way of determining $\beta$ in the limit of linear damping is by driving the spring-magnet system by a periodic emf at frequencies around the natural frequency ($\omega_0$) of the system to observe the phenomenon of resonance. As explained earlier, after a brief transient period, the driven oscillator achieves steady oscillations at exactly the driving frequency $\omega$, and the amplitude of oscillation $A(\omega)$ is maximum when the resonant frequency is reached. The resonant frequency under damping is always less than $\omega_0$ because a damped oscillator moves slower than an undamped one thereby taking longer time to complete a cycle. The resonant frequency under viscous damping ($\omega_d$) and the undamped natural frequency are related by the equation $\omega_d^2 = \omega_0^2 - \beta^2$. In this case, we record the amplitude $\tilde A(\omega)$ of the oscillator at late times and confront it with the theoretical model response $A(\omega; \vec \lambda)$ given by Equation (\ref{eq:drivenOsc}), where $\vec \lambda \equiv \{F_0, \omega_0, \beta \}$ is a vector of (as yet) un-determined regression parameters. Assuming that the observational amplitude data $\tilde A$ are each independent, and corrupted by a zero-mean Gaussian distributed random error $\epsilon$, we have for the $i$th observation:
\begin{equation}
	|\tilde A(\omega_i)|^2 = |A(\omega_i; \vec \lambda)|^2 + \epsilon_i.
\end{equation}
The joint probability density of the model after $n$ observations is
\begin{equation}
	L(\vec \lambda, \sigma^2) \propto  \exp \left \{ \frac{- \sum_{i=1}^n \left ( |\tilde A(\omega_i)|^2 - |A(\omega_i; \vec \lambda)|^2 \right )^2}{2 \sigma^2} \right \}, 
\end{equation}
where $\sigma$ is the variance of the random errors at each observation. The joint probability density needs to be maximized by finding the minima of the sum of squared residuals $R(\vec \lambda) = \sum_{i=1}^n [ |\tilde A(\omega_i)|^2 - |A(\omega_i; \vec \lambda)|^2 ]^2 $. Operationally, this reduces to simultaneously solving a set of non-linear algebraic equations:  $\partial R (\vec \lambda) / \partial \vec \lambda = 0$ which gives the numerical values of $F_0, \omega_0$ and $\beta$. 

\begin{figure}
	\includegraphics[width=0.99\columnwidth]{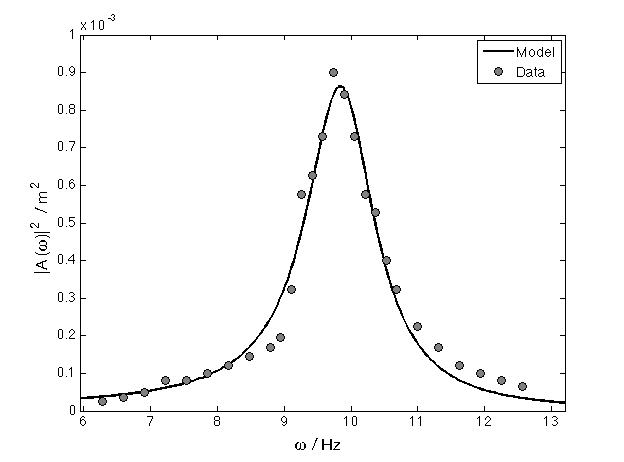}
	\caption{Resonance of the driven EMR apparatus in water (with linear viscous damping). As explained in the text, under the action of a periodic driving force of frequency $\omega$, the late time squared amplitude response has a Lorentzian shape. The model parameters are determined by non linear regression. If there are non-linear damping terms, the resonance curve deviates significantly from this generic shape.}
	\label{fig:resonanceInWater}
\end{figure}

In Figure \ref{fig:resonanceInWater} we show the squared amplitude response of the driven oscillator around resonant frequency and the {\em best fit} model which minimizes the squared residuals as explained before. The corresponding values of the model parameters are determined to be:  $F_0 = 0.36 \; [0.33, 0.39]\; {\mathrm{N/kg}}$, $(\omega_0/2 \pi) = 1.573 \; [1.566, 1.579] \; {\mathrm{Hz}}$ and $\beta = 0.623 \; [0.560, 0.687]\; {\mathrm{Hz}}$, where the numbers in brackets denote the 95\% confidence intervals of the respective quantities.  

A few comments are in order: first, the values obtained for the model parameters using non-linear regression in the driven-oscillator case are rather sensitive to the choice of initial parameters and an educated guess is required to obtain a sensible result. Secondly, since the value of the natural frequency $\omega_0$ is already known, one can in principle have a simpler model with only two unknown parameters $\{F_0, \beta\}$. Finally, the value of $\beta$ obtained by the two methods agree quite well within error bars - lending credence to both. This provides for independent measurements of the damping coefficient of water. Recall the fact that in the (strict) limit of small velocities and laminar flow, $\beta$ is linearly related to the dynamic co-efficient of viscosity $\eta$ modulo geometric factors of the magnet immersed \cite{Leith}. For a spherically shaped magnet of radius $R$, $\beta = (3 \pi R /m) \eta$. Using this relation, one can  calculate the numerical value of $\eta$:  we find that the value of $\eta$ thus obtained is two orders of magnitude larger than standard published value \citep{etaWater}. This is not surprising, given the fact that we are not operating in the limit that the linear relationship between $\beta$ and $\eta$ holds. Even so, $\beta$ {\em is} a measure of the magnitude of the viscous forces acting on the magnet which is accurately measured by our apparatus. 

\section{Comments and suggestions}
\label{sec:Discussion}

The design of an electromagnetically coupled oscillator is presented that can be used to demonstrate key concepts in a Waves \& Oscillations course at undergraduate level. Traditionally, many key concepts in such courses appear as solutions to second order ordinary differential equations and such a device can be used to play with the parameters corresponding to different terms of ODEs and provides a low-cost platform for a richer pedagogic experience to supplement classroom instructions. We suggest some activities for students to explore this setup further. We show that with minimal modifications, the same device is able to measure the linear viscous damping forces in a liquid: both by studying the decay amplitude of the damped system oscillating in the liquid and also by observing the resonance phenomenon when driven by a periodic external force. In both these cases, we find that the estimated value of the damping coefficient $\beta$ (which measures the damping force) agree within experimental errors. Students can explore the temperature dependence of $\beta$ with a help of a PID temperature controller. Coils of more turns can be used to try and enhance the induced emf thereby improving the sensitivity of the instrument. Viscous forces due to binary liquids may also be calculated which in turn may lead to finding the concentration of the solute and the solvent \citep{Rani}. Thus a wide-range  of experimental activities can be designed around such a device. 

\begin{figure}
	\includegraphics[width=0.99\columnwidth]{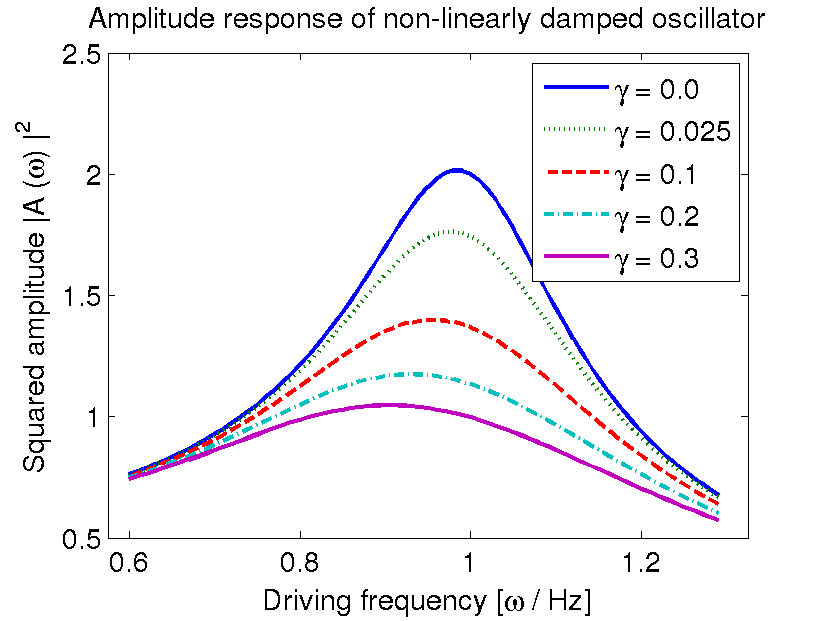}
	\caption{Late time amplitude response (squared) $|A(\omega)|^2$ of a driven, non-linearly damped oscillator as described by the equation: $\ddot z + \kappa \dot z + {\mathrm{sgn}}(\dot z) \gamma \dot z ^2 + \Omega z = {\mathcal A} \sin (\omega t)$ for fiducial values of $\kappa = 0.25, \; \Omega = 1.0 \; {\mathcal A} = 1.0$ and different values of $\gamma$. Note that in this case, the coefficient $\kappa$ and $\gamma$ are measures of linear and non-linear damping respectively. From the discussion in Section \ref{sec:Discussion}, one can identify $\kappa = b_1/m$ and $\gamma = b_2/m$, where $m$ is the mass of the system. While this equation is hard to solve analytically to provide a model (analytical) amplitude response to be compared against experiments, we provide some ideas in the text for probing the non-linear terms in damping. Note that as the non-linear damping term gets stronger, the shape of the amplitude response changes from the Lorentzian shape it has in the case of $\gamma = 0$ (linear-damping only).}  
	\label{fig:forcedDampingNumerical}
\end{figure}

Although we have assumed linear drag forces earlier, such an apparatus may also enable students to explore higher order viscous forces acting on the magnet. To next order in velocity, the total drag force acting on the oscillating magnet is given by : $-b_1 v - {\mathrm {sgn}}(v) \; b_2 v^2$ where $b_1$ and $b_2$ are constants and $v$ is the instantaneous velocity of the magnet. While it is  difficult to solve the ODE describing such a system,  one expects intuitively that the additional viscous term should dissipate energy faster thus accelerating the damping of amplitude. In other words, the quality factor of such oscillators should be lower. One may try to solve the ODE perturbatively (assuming $b_2 \ll b_1$) to arrive at a model solution against which the observation can be confronted to estimate $b_2$. However, numerical solutions to such ODEs show that the envelope of the decaying amplitude does not vary much in shape to allow for parameter estimation of non-linear terms. When such systems are driven externally, a numerical investigation (Figure \ref{fig:forcedDampingNumerical}) reveals that the shape of the resonance (peak and width) considerably changes by varying $b_2$. This is encouraging as it might be easier to separate the damping effects at linear and non-linear orders and ODE regression methods \cite{Brunel} may be explored in such cases to determine the damping constants. We also suggest that students explore the possibility of adding a phenomenological term in the denominator of Equation (\ref{eq:drivenOsc}) to model the oscillations in this case and follow the method of non-linear regression outlined earlier to determine the parameters. Recently, a method of directly checking the differential equation of the motion without integrating it has been suggested \cite{Moreno} using video photogrammetry. This can be very useful in analyzing such non-linear oscillations.

\subsection*{Acknowledgements}
We thank Prof. Sudhir K. Jain (Director, IIT Gandhinagar) for support and funding. We also thank Dr. S. Sarkar and Dr. S. Jolad for many illuminating discussions, fellow sophomores Mukesh Kumar, Pradeep Diwakar, Naman Singh and Naveen Kumar for their help with the experiment. Last but certainly not the least, we thank Mr. Mayur Chauhan in the UG Physics laboratory for helping us build the apparatus.

\end{document}